\documentclass[aip,amsmath,amssymb,reprint]{revtex4-1}

\usepackage{graphicx} % with figures

\usepackage{color,soul} % enable \hl{}

\usepackage{dcolumn}% Align table columns on decimal point
\usepackage{bm}% bold math
\usepackage[utf8]{inputenc}
\usepackage[T1]{fontenc}
\usepackage{mathptmx}

\draft % marks overfull lines with a black rule on the right

\begin{document}

\title{Microwave filtering using forward Brillouin scattering in photonic-phononic emit-receive devices} %Title of paper

\author{Shai Gertler}
\email[]{shai.gertler@yale.edu}
\affiliation{Department of Applied Physics, Yale University, New Haven, CT 06520, USA}

\author{Eric A. Kittlaus}
\affiliation{Department of Applied Physics, Yale University, New Haven, CT 06520, USA}
\affiliation{Jet Propulsion Laboratory, California Institute of Technology, Pasadena, CA 91109, USA}

\author{Nils T. Otterstrom}
\affiliation{Department of Applied Physics, Yale University, New Haven, CT 06520, USA}

\author{Prashanta Kharel}
\affiliation{Department of Applied Physics, Yale University, New Haven, CT 06520, USA}

\author{Peter T. Rakich}
\email[]{peter.rakich@yale.edu}
\affiliation{Department of Applied Physics, Yale University, New Haven, CT 06520, USA}

% \date{\today}

\begin{abstract}
Microwave photonic systems are compelling for their ability to process signals at high frequencies and over extremely wide bandwidths as a basis for next generation communication and radar technologies. 
However, many applications also require narrow-band $(\sim\text{MHz})$ filtering operations that are challenging to implement using optical filtering techniques, as this requires reliable integration of ultra-high quality factor $(\sim 10^8)$ optical resonators.
One way to address this challenge is to utilize long-lived acoustic resonances, taking advantage of their narrow-band frequency response to filter microwave signals. 
In this paper, we examine new strategies to harness a narrow-band acoustic response within a microwave-photonic signal processing platform through the use of light-sound coupling. 
Our signal processing scheme is based on a recently demonstrated photon-phonon emitter-receiver device, which transfers information between the optical and acoustic domains using Brillouin interactions, and produces narrow-band filtering of a microwave signal. 
To understand the best way to use this device technology, we study the properties of a microwave-photonic link using this filtering scheme. We theoretically analyze the noise characteristics of this microwave-photonic link, and explore the parameter space for the design and optimization of such systems.
\end{abstract}

\pacs{}% insert suggested PACS numbers in braces on next line

\maketitle %\maketitle must follow title, authors, abstract and \pacs

% Body of paper goes here. Use proper sectioning commands. \section{\label{sec:intro}Introduction}
As we seek to utilize an ever-increasing portion of the electromagnetic spectrum for next generation communications and radar systems, microwave-photonic signal processing platforms show great promise for their ability to manipulate signals at high frequencies and over extremely wide bandwidths \cite{capmany2006tutorial,marpaung2015low,marpaung2019integrated}. Rapid progress in the field of silicon photonics has also enabled the integration of high-speed modulators \cite{wang2018integrated,sun2018mo}, amplifiers \cite{liu2006nonlinear,foster2006broad,kittlaus2016_FSBS} and detectors \cite{baehr2005optical} with electronic and photonic circuitry \cite{chen2018emergence,soref2006past,izhaky2006development,atabaki2018integrating}, opening the door to miniaturization of radio-frequency (RF) photonic circuits having signal processing performance that is competitive with established microwave technologies \cite{Jalali2006_Silicon_Photonics_review,capmany2007microwave}. However, it remains challenging to implement narrow-band filtering using all-optical techniques, as ultra-low loss waveguides are needed to store signals for long periods of time \cite{dong2010ghz,tien2011ultra,dai2012passive}.

Narrow-band filtering operations can also be realized by accessing long-lived acoustic phonons through stimulated Brillouin scattering (SBS), a nonlinear three-wave interaction that produces coupling between optical waves and GHz-frequency elastic waves. While Brillouin interactions have been used to implement an array of filtering and delay operations \cite{zhu2007stored,morrison2014tunable,marpaung2015low,liu2016lossless,choudhary2018chip,merklein2018brillouin,xie2019system}, in this work we focus on bandpass filtering. 
Brillouin-based bandpass filtering operations are conventionally achieved by making use of the narrow-band optical amplification supplied by the stimulated Brillouin scattering process \cite{byrnes2012photonic,stern2014tunable}.
While such amplification-based filtering strategies have been used to synthesize highly desirable response functions \cite{morrison2014tunable,marpaung2015low,liu2016lossless}, the high levels of Brillouin gain necessary to implement these schemes also enhance unwanted noise sources that can degrade the performance of the RF link \cite{xie2019system}.
Other filtering schemes have demonstrated pass-band filtering by utilizing Brillouin-induced loss on both sides of the desired pass-band, achieving a lower noise-figure at the expense of the filter bandwidth, out-of-band rejection, and stop-band range \cite{xie2019system}.
In many cases, the noise-figure, dynamic range, and out-of-band rejection of such filters do not meet the increasingly stringent requirements of numerous applications \cite{liu2018high,marpaung2019integrated}.

Alternatively, Brillouin interactions can produce narrowband filtering without using an amplification process through use of a recently demonstrated photonic-phononic emit-receive (PPER) device design. Within this PPER device, the signal is converted to an acoustic wave that transfers information between two spatially distinct waveguides over a narrow spectral band. The transfer function of a PPER filter is synthesized by tailoring the acoustic-wave response to produce narrowband (3 MHz) multi-pole filters with high (70 dB) out-of-band rejection \cite{shin2015control}. The spatial separation of the `emit' and `receive' optical waves decouples the input signal from spontaneous scattering, which is a fundamental noise source in Brillouin-based devices. This yields a different design space from other Brillouin-based RF-photonic filters, which may prove advantageous for many practical applications. 
The recent demonstrations of PPER were done in silicon and silicon-nitride platforms, where suspended waveguides with micron-scale cross sections were used to guide both optical and acoustic waves \cite{shin2015control,kittlaus2018rf,kittlaus2017_Intermodal}. The forward Brillouin scattering process used in these devices permits tailoring of the acoustic frequency over a large range by modifying the dimensions of the acoustic waveguide \cite{shin2013tailorable}.
A proof-of-principle RF-photonic link was demonstrated using such PPER devices, revealing a promising RF link gain ($-2$ dB) and spurious-free dynamic range (SFDR) (99 dB Hz$^{2/3}$) \cite{kittlaus2018rf}. In order to improve upon these results, it is necessary to identify system parameters that will provide the greatest opportunity to achieve high performance.

In this paper, we present a systematic analysis of a microwave-photonic filter designed around a PPER device, and identify the key characteristics of the system that dictate the link performance. Our model of the system includes the effects of noise associated with the laser sources and optical detection, as well as excess noise produced by spontaneous Brillouin scattering within the PPER device. Our analysis reveals that the spontaneous Brillouin scattering from thermally-populated acoustic phonons is the dominant noise source within the system over a range of operating conditions. Nevertheless, we show that this strategy holds the potential for high performance based on commercially available modulators and detector technologies. Furthermore, with the development of new electro-optic modulator technologies \cite{van2009low,wang2018integrated} and high power-handling capabilities \cite{rong2005continuous,tien2011ultra,DARPAridgway2014microwave}, we show how PPER based filters with a high RF-link gain ($>$45 dB), a large dynamic range ($>$110 dB Hz$^{2/3}$) and a low noise figure ($<$10 dB) are possible.
%%%
\section{\label{sec:theory}Theoretical study}
\begin{figure*}
    \centering
    \includegraphics[scale=0.8]{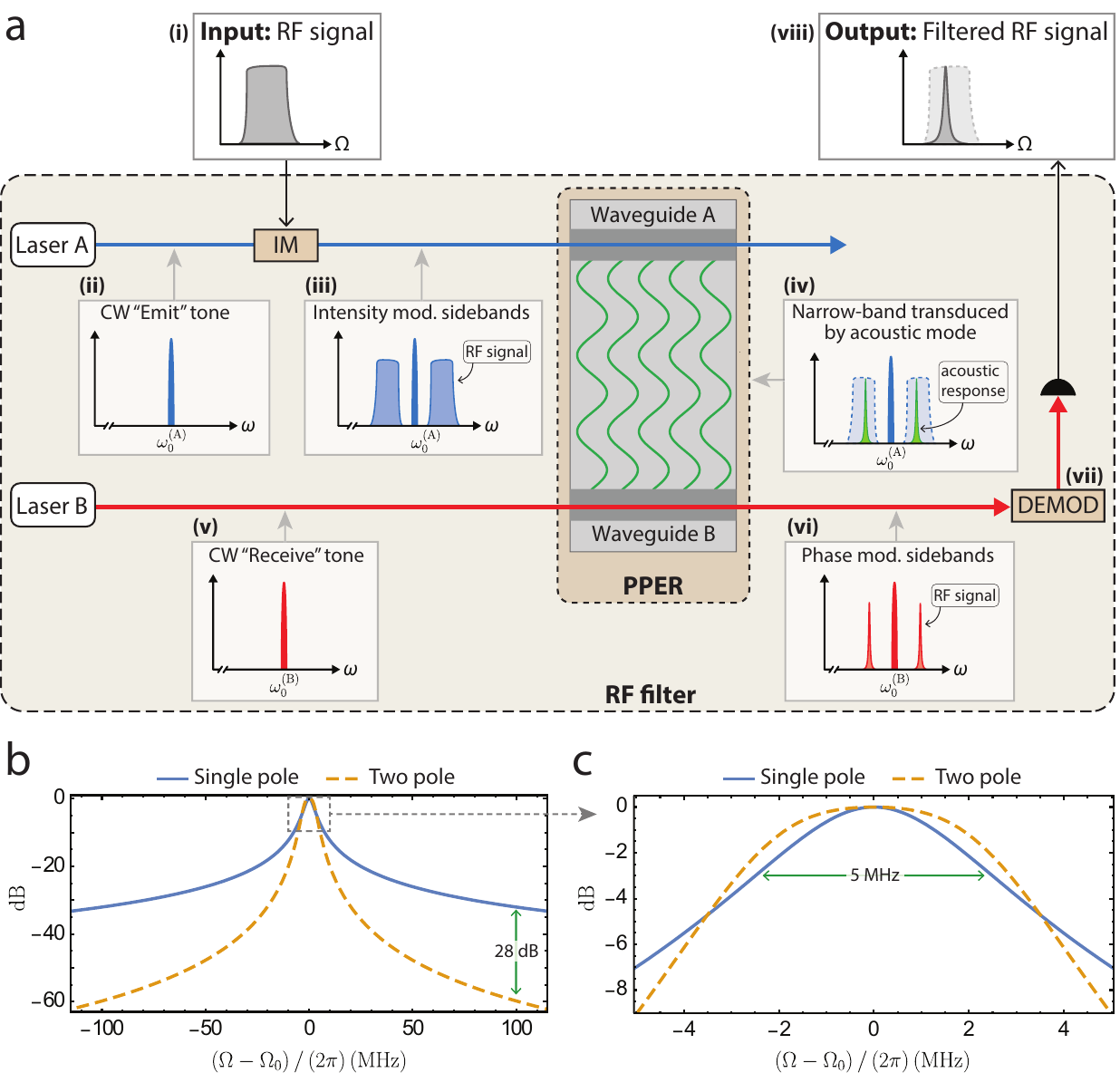}
    \caption{
    a. Operation scheme of a microwave filter designed around a PPER device. The RF input (i) is used to modulate a laser tone (ii), such that the RF information is in the optical sidebands (iii) and directed into waveguide A of the PPER. A narrow-band acoustic field (iv) mediates the information to waveguide B, and modulates a separate optical tone (v) in the form of phase modulation (vi). The phase modulation sidebands are demodulated (vii) and the filtered RF information is retrieved at the filter output (viii). IM: intensity modulator, DEMOD: phase demodulation.
    b. Normalized frequency response of a PPER filter, for both a single and two-pole design. The two-pole filter yields a sharper frequency roll-off, with an improvement of 28 dB out-of-band rejection 100 MHz from the pass-band center frequency.
    c. Magnified view of the pass-band center shows a flat-top frequency response, closer to an ideal band-pass filter. The plots shown were calculated using the filter parameters from Table \ref{tbl:params}.
    }
    \label{fig:RF_system}
\end{figure*}

In what follows, we explore the properties of a narrow-band microwave photonic filter based on a PPER device of the type described above.
Throughout this paper, we consider the operation of a PPER device within a microwave-photonic link of the type seen in the block diagram of Fig. \ref{fig:RF_system}(a). This system is used to filter a wideband RF signal that enters through the RF input port and exits at the RF output port after taking on the spectral characteristics produced by the acoustic transfer function of the PPER device.

We begin by describing how the input RF signal is shaped as it is converted between the microwave, optical, and acoustic domains when passing through the system depicted in Fig. \ref{fig:RF_system}(a). The incident RF signal, seen in panel (i), is encoded on an optical carrier produced by Laser A, panel (ii), using an intensity modulator (IM), as shown in panel (iii). The modulated field is injected into the emit waveguide of the device, denoted waveguide A. As the laser field traverses the device, a portion of the signal wave is transduced as gigahertz-frequency acoustic waves (green) over a narrow spectral band, determined by the device geometry \cite{rakich2010tailoring,rakich2012giant,shin2013tailorable}, as seen in panel (iv). A second laser field produced by Laser B, seen in panel (v) is injected into a spatially-separated receive waveguide (waveguide B) that is used to sense the transduced acoustic wave. This spectrally-filtered replica of the microwave signal is encoded on the light propagating in the receive waveguide as pure phase modulation, illustrated in panel (vi). The phase-modulated signal wave exiting the receive waveguide is then passed through a demodulator (vii) that converts phase modulation into intensity modulation and is converted back to the microwave domain at the output (viii). The resulting output RF signal is filtered, taking on the shape of the acoustic transfer function.

Next, we outline the analytical form of the signal wave as it traverses the photonic filter, which allows us to identify key parameters of the system. At the microwave input, the input RF voltage with amplitude $V_\text{in}$ and frequency $\Omega$ modulates an optical carrier with power $P^\text{(A)}$ and optical frequency $\omega_0^\text{(A)}$. The intensity-modulated optical field can be described by \cite{urick2015fundamentals}
\begin{equation}
    E^\text{(A)}_\text{in}(t) = \sqrt{P^\text{(A)}} e^{-i \omega_0^\text{(A)} t} \left(\frac{1}{\sqrt{2}} e^{i \frac{\pi}{2}} + \frac{1}{\sqrt{2}} e^{-i \pi\frac{V_\text{in}}{V_\pi} \sin{\left(\Omega t\right)} }
    \right),
    \label{eq:IM_input}
\end{equation}
where \(V_\pi \) is the half-wave voltage of the modulator. This field is then directed into waveguide A, where it drives a coherent acoustic field through forward Brillouin scattering \cite{shin2013tailorable}, with a displacement amplitude determined by the acoustic resonance properties, the Brillouin coupling rate, and the input optical power. A second optical source is directed into waveguide B with optical power $P^\text{(B)}$ and optical frequency $\omega_0^\text{(B)}$. Propagating through the device, the light is phase modulated by the driven acoustic field, and after a distance $L$, at the output of waveguide B, the field can be described by
\begin{equation}
    E^\text{(B)}_\text{out}(t) = \sqrt{P^\text{(B)}} \ e^{-i \left(\omega^\text{(B)}_0 t - \beta_\text{in} \cos{\left(\Omega t - \phi\right)}\right)},
    \label{eq:phase_mod_out}
\end{equation} 
showing a pure phase modulation of the optical tone. The modulation index is determined by the system and device parameters, given by 
\begin{equation}
\beta_\text{in} =\frac{\Gamma}{2} \left|\chi(\Omega)\right| J_1\left(\pi\frac{V_\text{in}}{V_{\pi}}\right)G_\text{B} P^\text{(A)} L,
\label{eq:phase_mod_index}
\end{equation}
where \(\Gamma\) is the acoustic dissipation rate, \(\chi(\Omega)\) is the frequency response of the acoustic resonance, and \(J_1(\cdot)\) denotes a Bessel function of first order, describing the power in the sidebands of the intensity modulated input field. \(G_\text{B}\) is the Brillouin gain, a metric for the strength of the light-sound interaction, and \(L\) the length of the region where the Brillouin coupling takes place. The phase of this response is denoted \(\phi = \arg{\left( \chi \right)}\), giving an overall phase shift to the phase modulation, as seen in Eq. (\ref{eq:phase_mod_out}). More details can be found in Appendix \ref{subsec:app_IM_input}.

While numerous schemes can be used to demodulate this phase modulated signal \cite{zhang2008phase,zibar2008phase,urick2015fundamentals}, for simplicity we consider the use of an imbalanced Mach-Zehnder interferometer (MZI). Assuming that the MZI is biased in quadrature followed by photodiodes with responsivity $\eta$, the RF power at the link output is given by
\begin{widetext}
\begin{equation}
    P_\text{out}^\text{RF}(\Omega) = 2 \Bigg(\eta P^\text{(B)} J_1 \left( \Gamma \left|\chi(\Omega)\right| J_1\left(\pi\frac{V_\text{in}}{V_{\pi}}\right)G_\text{B} P^\text{(A)} L \sin{(\Omega \tau/2)}\right) \Bigg)^2 R_\text{out} |H_\text{pd}|^2,
    \label{eq:P_out_RF}
\end{equation}
\end{widetext}
where $R_\text{out}$ is the output impedance, $H_\text{pd}$ is the photodiode circuit efficiency \cite{urick2015fundamentals}, and $\tau$ is the relative time delay between the two arms of the MZI. A detailed derivation can be found in Appendix \ref{subsec:app_bal_demod}. These results can easily be modified for different demodulation schemes, and an example of an alternative approach using an optical notch filter is described in Appendix \ref{subsec:filt_demod}.

The frequency response of the filter $\chi(\Omega)$ is shaped by the acoustic resonances taking part in the signal transduction, and can be tailored through the geometry of the device and acoustic mode engineering \cite{shin2013tailorable,shin2015control,OTHER_PAPER}. A single acoustic mode will result in a Lorentzian line shape, given by
$ \chi(\Omega) = \left[i(\Omega_0-\Omega)+\left(\Gamma/2\right)\right]^{-1} $,
determined by the acoustic resonant frequency $\Omega_0$ and dissipation rate $\Gamma$, as detailed in Appendix \ref{subsec:app_IM_input}. Such a single pole filter was recently demonstrated, with a pass-band frequency of 4.3 GHz and a 5 MHz linewidth \cite{kittlaus2018rf}. However, the PPER design is not limited to this single-pole frequency response. A PPER scheme can be designed to exhibit a multi-pole frequency response by using multiple acoustic modes in the filtering process. For example, in the case of two identical acoustic modes with a resonant frequency $\Omega_0$ and a coupling rate $\mu$, the frequency response is given by \cite{shin2015control,OTHER_PAPER}
$
    \chi(\Omega)^\text{(2 pole)} = \left(i \mu\right) \left[\left(i \left(\Omega_0 + \mu - \Omega\right) + \left(\Gamma/2\right)\right)\left(i \left(\Omega_0 - \mu - \Omega\right) + \left(\Gamma/2\right)\right)\right]^{-1}.
    \label{eq:chi_2pole}
$
This multi-pole response exhibits a sharp frequency roll-off, resulting in high out-of-band rejection, as demonstrated in Fig. \ref{fig:RF_system}(b), showing 60 dB of suppression for frequencies $>100$ MHz from the filter center frequency, an improvement of 28 dB compared to an equivalent single-pole filter. This property is of great importance for applications such as channelizers where different spectral bands may interfere in the absence of high out-of-band rejection \cite{anderson1991advanced}. Furthermore, the pass-band can be designed to have a flatter frequency response, as shown in Fig. \ref{fig:RF_system}(c), when comparing to a typical Lorentzian line-shape, yielding a closer approximation of an ideal band-pass filter. A second-order PPER based filter has been demonstrated, showing a 3.15 MHz pass-band, and 70 dB rejection 100 MHz from the center frequency \cite{shin2015control}. This approach can be further extended to higher-order filter responses, by coupling a larger number of acoustic resonators \cite{OTHER_PAPER}.

\subsection*{\label{subsec:noise}Noise sources}
A PPER device utilizes the optical, acoustic and RF domains to implement filtering. Hence, for a full description of the device properties, noise sources from all the different domains need to be considered.
Following the standard convention in RF photonics {\cite{urick2015fundamentals}}, we express the noise power per unit frequency as
\begin{eqnarray}
    && N_\text{out} =  k_\text{B} T + g k_\text{B} T +2 q \eta P^{(B)} R_\text{out} |H_\text{pd}|^2 + \nonumber\\
    && \left(\eta P^{(B)} \right)^2 R_\text{out} |H_\text{pd}|^2 \Big(\text{RIN} + \text{RIN}_\text{phase} + \text{RIN}_\text{phonons} \Big).
    \label{eq:all_noise}
\end{eqnarray}

This formulation captures noise contributions from the optical, microwave, and acoustic domains.
The first term in Eq. (\ref{eq:all_noise}) is the RF thermal noise at the detector, where $T$ denotes the temperature and $k_\text{B}$ is the Boltzmann constant. The second term is the thermal RF noise at the link input going through the filter with an RF link-gain $g$, analyzed in the next section. The third term is due to shot noise, where $q$ is the electron charge, and scales linearly with the optical power incident on the detector. The first relative intensity noise (RIN) term accounts for intensity fluctuations of the optical sources, which are usually negligible at frequencies over a few GHz. The next term accounts for the demodulation at the device output turning phase fluctuations into intensity, such that phase noise from the laser source is converted to intensity noise. Assuming the phase noise has a Lorentzian spectral line-shape with full width half maximum \(\gamma\), it has been shown that detection using a MZI will result in a noise power spectral density given by \cite{moslehi1986noise,tkach1986phase,urick2015fundamentals}
\begin{equation}
    \text{RIN}_\text{phase}^\text{bal} = e^{-\gamma \tau} \left(\frac{2 \gamma}{\gamma^2 + \Omega^2}\right) \Big( \cosh{(\gamma \tau)} -\cos{(\Omega \tau)} \Big).
\end{equation}

The last term in Eq. (\ref{eq:all_noise}) is the result of the thermal occupation of the acoustic modes taking part in the Brillouin scattering process. At non-zero temperature, thermally driven fluctuations will add phase noise to the optical field in the receive waveguide, which can also be described in terms of spontaneous forward-Brillouin scattering \cite{boyd_book,kharel2016_Hamiltonian}. The phase demodulation will convert this into intensity fluctuations, and the RIN associated with this noise source, detailed in Appendix \ref{subsec:th_ph_noise}, is given by 
\begin{equation}
    \text{RIN}_\text{phonons}^\text{bal} = 8 \frac{\omega_0}{\Omega_0} G_\text{B} L k_\text{B} T \sin^2{\left( \Omega \tau /2 \right)} \left(\frac{\Gamma}{2}\right)^2 \left| \chi_N (\Omega)\right|^2,
    \label{eq:thermal_ph_bal}
\end{equation}
where \(T\) is the temperature, \(k_\text{B}\) the Boltzmann constant, and \(\chi_N(\Omega)\) the frequency spectrum of the spontaneous Brillouin scattering. For a single acoustic mode, the spontaneous scattering follows a Lorentzian lineshape \cite{kharel2016_Hamiltonian}, such that
$ | \chi_N |^2 = \left[(\Omega - \Omega_0)^2+(\Gamma/2)^2\right]^{-1} $.
When analyzing high-order filters, all of the acoustic modes taking part in the filtering process contribute to the spontaneous scattering \cite{OTHER_PAPER}. In the case of two identical acoustic modes with a coupling rate $\mu$, this will yield 
$ | \chi_N^\text{(2 pole)} |^2 = (1/2) ([(\Omega - \Omega_0 + \mu)^2+(\Gamma/2)^2]^{-1} + [(\Omega - \Omega_0 - \mu)^2+(\Gamma/2)^2]^{-1} )$,
which exhibits a modified frequency response as a result of the two resonances, but decays as a single pole Lorentzian away from the pass-band.

While spontaneous Brillouin scattering does occur in the emit waveguide, it does not degrade the signal-to-noise of the system. This is because only intensity modulated light fields contribute to the coherent transduction of information, while the spontaneous scattering results only in phase modulation, independent of the driving field in the emit waveguide. 
As illustrated in Fig. \ref{fig:RF_properties}(c), the dominant noise contribution in a PPER based filter at room temperature is the thermal Brillouin scattering, and this has also been demonstrated experimentally \cite{kittlaus2018rf}. This noise is centered around the center frequency, and does not contribute excess noise out of band.

\subsection*{\label{subsec:RF_link}RF link properties}
Analyzing the performance metrics of a RF-photonic link is important for practical deployment and interfacing it with other components in a microwave system. In order to calculate useful RF link properties, we start by looking at the linear small-signal gain of the PPER filter. We express the output RF power in terms of the input RF power $P_\text{in}^{\text{RF}}=V_{\text{in}}^2/(2R_{\text{in}})$
\cite{urick2015fundamentals}, and analyze the linear operating region of the filter by expanding the Bessel functions in Eq. (\ref{eq:P_out_RF}) to first order. This yields a linear transfer function between the input and output RF power ($g=P_\text{out}^\text{RF}/P_\text{in}^\text{RF}$), defined as the RF link-gain
\begin{equation}
    g = R_\text{in}R_\text{out} |H_\text{pd}|^2 \Bigg(\eta P^\text{(B)} \frac{\Gamma}{2} \left|\chi(\Omega)\right| \frac{\pi}{V_{\pi}} G_\text{B} P^\text{(A)} L \sin{(\Omega \tau/2)}\Bigg)^2,
    \label{eq:P_out_RF_lin2}
\end{equation}
where $ R_{\text{in}} $ is the input impedance of the intensity modulator. Eq. (\ref{eq:P_out_RF_lin2}) shows how the link-gain improves quadratically with the optical powers, Brillouin gain, and device length. We can also see that for maximum signal output, we need the time delay of the phase-demodulation MZI to satisfy $\Omega_0\tau/2 = \pi (m+1/2) $.

\begin{figure*}
    \centering
    \includegraphics[scale=0.8]{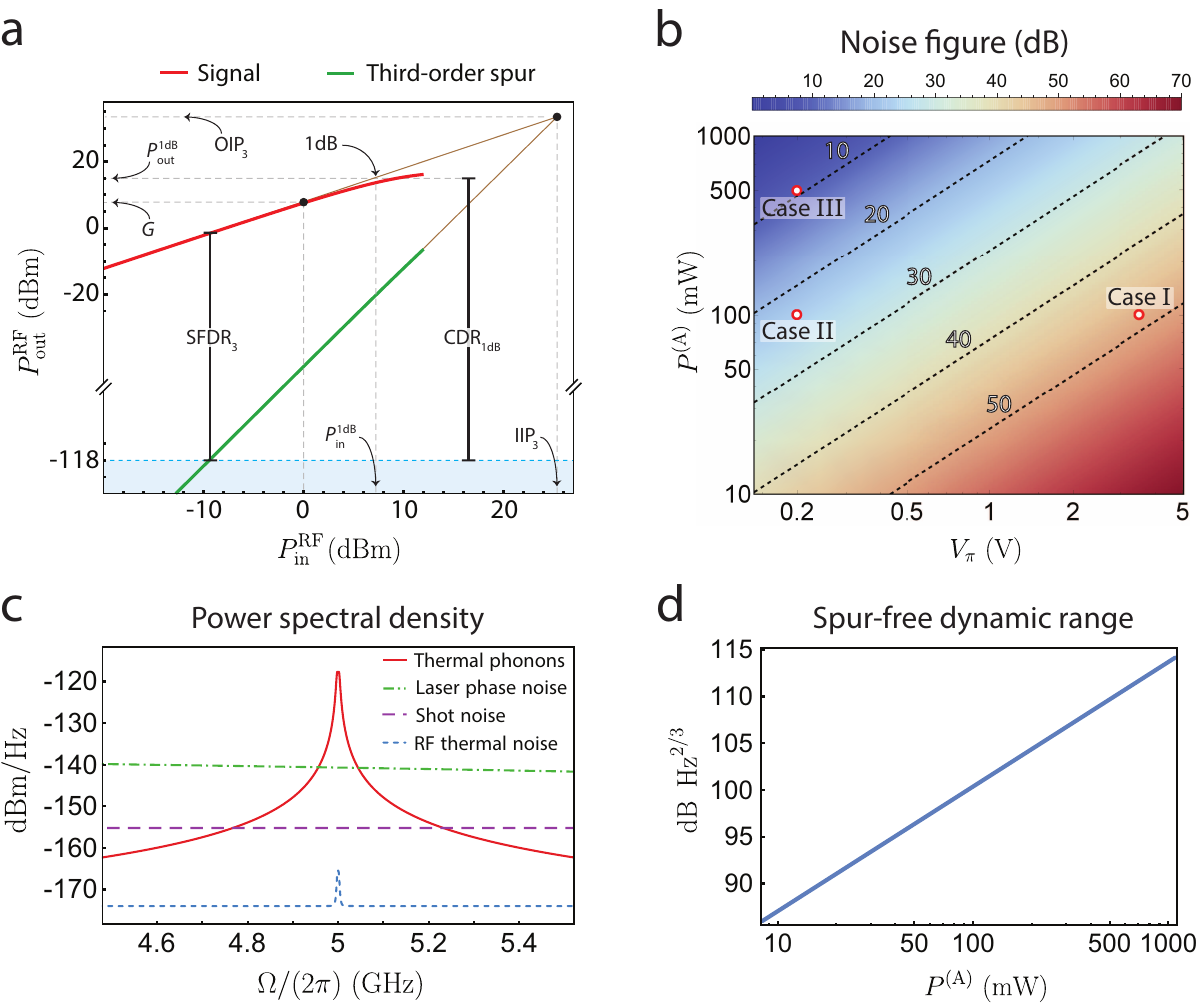}
    \caption{
    a. Calculated RF output power as a function of RF input power using the parameters from Table \ref{tbl:params}, illustrating the signal (red), the third-order spur (green), and the link noise-floor (blue). 
    b. Contour plot of the calculated noise figure of a PPER based RF link, as a function of optical power in waveguide A and the intensity modulator $V_\pi$, using the parameters detailed in Table \ref{tbl:params}. The three cases detailed in Table \ref{tbl:RF_fig_merit} are shown for reference.
    c. Calculated noise sources in a PPER RF-photonic link, using the device parameters detailed in Table \ref{tbl:params}.
    d. Calculated $\text{SFDR}_3$ as a function of optical power in waveguide A, using $V_\pi = 3.5$ V, and the parameters detailed in Table \ref{tbl:params}.
    }
    \label{fig:RF_properties}
\end{figure*}

Next, we analyze the noise floor of the RF link, assuming the thermal phonon noise is the dominant noise source. We use Eqs. (\ref{eq:all_noise}) and (\ref{eq:thermal_ph_bal}), and calculate the noise power in a  frequency band $B_\text{RF}$, much narrower relative to the filter bandwidth, yielding
\begin{eqnarray}
    P_\text{N}^\text{bal}(\Omega) &&= 8 \left(\eta P^\text{(B) } \sin{\left( \Omega \tau /2 \right)} \right)^2 \frac{\omega_0}{\Omega_0} \nonumber\\
    &&\times G_\text{B} L k_\text{B} T B_\text{RF} R_\text{out} |H_\text{pd}|^2 \left(\frac{\Gamma}{2}\right)^2 \left| \chi_N (\Omega)\right|^2.
    \label{eq:PN_bal}
\end{eqnarray}

To assess the non-linearity of the device, we first calculate the \(1 \ \text{dB} \) compression point, i.e. the RF input power at which the output power is 1 dB lower than the linear response predicts. Setting the ratio between Eqs. (\ref{eq:P_out_RF}) and (\ref{eq:P_out_RF_lin2}) to 1 dB (\(10^{0.1}\)), we can solve numerically to find the input RF power
\begin{equation}
    P_\text{in}^{\text{1dB}} =\frac{0.183}{R_\text{in}} \left(\frac{V_{\pi}}{\Gamma \left|\chi(\Omega)\right| G_\text{B} P^\text{(A)} L \sin{(\Omega \tau/2)}}\right)^2,
    \label{eq:Pin1dB}
\end{equation}
and the linear dynamic range, defined as $\text{CDR}_{1\text{dB}} = g P_\text{in}^{\text{1dB}} / P_\text{N}$ and illustrated visually in Fig. \ref{fig:RF_properties}(a), can be expressed as
\begin{equation}
    \text{CDR}_{1\text{dB}} = 0.226 \frac{\Omega_0}{\omega_0^\text{(B)}} \frac{1}{k_\text{B} T B_\text{RF}} \frac{1}{\Gamma^2 \left|\chi(\Omega)\right|^2 G_\text{B} L \sin^2{(\Omega \tau/2)}}.
    \label{eq:CDR}
\end{equation}
In this calculation we have assumed that the compression is a result of the intensity modulator at the link input, consistent with parameters of recently demonstrated devices. A similar expression can be derived for the distortion resulting from the Brillouin process in the case of high nonlinear gain.

Another important measure of the performance of an RF filter is the third-order intercept point, denoting the power at which the linear signal is equal to the third-order spurious tone \cite{urick2015fundamentals}, illustrated in Fig. \ref{fig:RF_properties}(a). In the PPER  system, the spurious tone is a result of the intensity modulator not having a perfectly linear response, resulting in unwanted spurious signals. To quantify this, we assume the input RF signal is at frequency $\Omega/3$ and see the propagation of the third-order modulation through the device. In this case, Eq. (\ref{eq:phase_mod_index}) transforms into 
$
\beta_\text{in}^{3\Omega/3} = \left(\Gamma/2\right) \left|\chi(\Omega)\right| J_3\left(\pi V_\text{in}/V_{\pi}\right)G_\text{B} P_0^{(A)} L,
$
where the Bessel function is of order three as we are now interested in the third harmonic. Plugging into Eq. (\ref{eq:P_out_RF}) and expanding the Bessel functions to the first non-vanishing order we have
\begin{eqnarray}
&&P_\text{out}^{3\Omega/3} =\frac{1}{144}\left (P_\text{in}^{\text{RF}}\right)^3 R_\text{in}^3 R_\text{out} |H_\text{pd}|^2 \nonumber\\
&&\times \left(\eta P^\text{(B)}\frac{\Gamma}{2} \left|\chi(\Omega)\right|\left(\frac{\pi}{V_{\pi}}\right)^3 G_\text{B} P^\text{(A)} L \sin{(\Omega \tau/2)} \right)^2.
\label{eq:P_RF_out_3Omega_ss2}
\end{eqnarray}

To find the third-order intercept point, we equate Eqs. (\ref{eq:P_out_RF_lin2}) and (\ref{eq:P_RF_out_3Omega_ss2}), and solve for the input RF power, giving the input intercept point ($\text{IIP}_3$)
\begin{equation}
    \text{IIP}_3 = 12\frac{V_{\pi}^2}{\pi^2 R_\text{in}},
\end{equation}
and plugging back into Eq. (\ref{eq:P_out_RF_lin2}) gives us the output intercept point ($\text{OIP}_3$)
\begin{equation}
    \text{OIP}_3 = 12 \left(\eta P^\text{(B)} \frac{\Gamma}{2} \left|\chi(\Omega)\right| G_\text{B} P^\text{(A)} L \sin{(\Omega \tau/2)} \right)^2R_\text{out} |H_\text{pd}|^2.
    \label{eq:OIP3}
\end{equation}

We can now calculate the spurious free dynamic range, describing the range of RF power between the minimum detectable signal, up to appearance of the third-order spur, and given by $\text{SFDR}_3 = \left(\text{OIP}_3 / P_\text{N}\right)^{2/3}$. Using Eqs. (\ref{eq:OIP3}) and (\ref{eq:PN_bal}) we have
\begin{equation}
    \text{SFDR}_3 = \left(\frac{3}{2} \ \frac{\Omega_0}{\omega_0^\text{(B)}} \ \frac{G_\text{B} \left(P^\text{(A)}\right)^2 L}{  k_\text{B} T B_\text{RF}} \frac{\left| \chi (\Omega)\right|^2}{\left| \chi_N (\Omega)\right|^2} \right)^{2/3},
\end{equation}
showing higher dynamic range with higher optical power, device length and Brillouin gain.
We also look at the signal to noise ratio (SNR) for a given RF input signal with power $P_\text{in}^{\text{RF}}$
\begin{equation}
    \text{SNR} = \frac{1}{8} P_\text{in}^{\text{RF}} \left(\frac{\pi}{V_{\pi}}\right)^2 \frac{\Omega_0}{\omega_0^\text{(B)}} \frac{G_\text{B} \left(P^\text{(A)}\right)^2 L} {k_\text{B} T B_\text{RF}}  \frac{\left| \chi (\Omega) \right|^2}{\left| \chi_N (\Omega)\right|^2} R_\text{in},
    \label{eq:SNR}
\end{equation}
which can be improved by using a low $V_\pi$ modulator as well as high optical power.

An important metric for an RF link is the noise figure, quantifying the noise added by the RF link to an input signal, by analyzing the ratio of the input to output SNR, \(F=\text{SNR}_\text{in}/\text{SNR}_\text{out}\). Using the result from Eq. (\ref{eq:SNR}), and assuming thermal noise at the RF input port \(\left(k_\text{B} T B_\text{RF}\right)\) yields
\begin{equation}
    F = \frac{8}{R_\text{in}}\left(\frac{V_{\pi}}{\pi}\right)^2 \left(\frac{\omega_0^\text{(B)}}{\Omega_0}\right)\frac{1}{G_\text{B} \left(P^\text{(A)}\right)^2 L} \frac{\left| \chi_N (\Omega) \right|^2}{\left| \chi (\Omega) \right|^2}.
\end{equation}

Analyzing a two-pole filter, with the frequency responses of signal ($ \chi^\text{(2 pole)}$) and the noise ($ \chi_N^\text{(2 pole)}$), and using the parameters from Table \ref{tbl:params}, we calculate a noise figure of 48 dB and a spur-free dynamic range of 100 dB Hz$^{2/3}$. 
The system parameters chosen in our study are similar to those used in recent experiments \cite{shin2015control,kittlaus2018rf}, and the theoretical analysis is consistent with experimental results. For example, using the experimental parameters described in Ref. \cite{kittlaus2018rf} with no further calibration, we predict the measured gain, noise-figure and spur-free dynamic range to within a few dB.
As we seek higher performance systems of this type, it is instructive to explore a broader parameter space to inform future generations of PPER device design and RF links.  
Analyzing the design space of the system, illustrated in Figs. \ref{fig:RF_properties}(b) and \ref{fig:RF_properties}(d), reveals that we can further improve the performance of the RF link by using a low $V_\pi$ modulator and higher optical powers.
Lower $V_\pi$ corresponds to higher efficiency of the optical intensity modulation for the same input RF voltage \cite{bucholtz2008graphical}. This results in a stronger acoustic field transducing the RF signal, while not adding noise to the link.
We demonstrate the link parameter space by comparing link-parameters in three scenarios: recently demonstrated system (denoted Case \MakeUppercase{\romannumeral 1}), a PPER-based link employing a low half-wave voltage modulator (Case \MakeUppercase{\romannumeral 2}), and a link utilizing a low $V_\pi$ modulator as well as higher optical power (Case \MakeUppercase{\romannumeral 3}), as detailed in Table \ref{tbl:RF_fig_merit} and shown in Fig. \ref{fig:RF_properties}(b).
For example, using 500 mW in waveguide A and a modulator with $V_\pi = 0.2$ V will result in a noise figure of 9 dB, and dynamic range of 110 dB Hz$^{2/3}$. Additional improvement to the dynamic range can be achieved by using linearized modulators, as modulator-induced distortion is the limiting factor for linearity in this RF link \cite{bridges1995distortion,khilo2011broadband}.

\begin{table*}[!htbp]
\centering
\setlength{\tabcolsep}{8pt} % space between columns
\renewcommand{\arraystretch}{1.3} % space between rows
\begin{tabular}{| l | c | c | c | l |} \hline
Parameter & Case \MakeUppercase{\romannumeral 1} & Case \MakeUppercase{\romannumeral 2} & Case \MakeUppercase{\romannumeral 3} & Description  \\ \hline\hline
$G$ (dB) & 7.8 & 32.7 & 46.9 & RF link gain $\left(P_\text{out} / P_\text{in}\right)$  \\ \hline
$\text{OIP}_3$ (dBm) & 32.6 & 32.6 & 46.8 & Output intercept point  \\ \hline
$\text{SFDR}_3 \big(\text{dB Hz}^{2/3}\big)$ & 100.4 & 100.4 & 112.4 & Spur-free dynamic range $\left(\text{OIP}_3/P_\text{N}\right)^{2/3}$  \\ \hline
$P_\text{out}^{\text{1dB}}$ (dBm) & 15 & 15 & 10.3 & 1 dB compression point  \\ \hline
$\text{CDR}_{1\text{dB}}$ (dB Hz) & 133 & 133 & 132 & Linear dynamic range $\left( P_\text{out}^{\text{1dB}} / P_\text{N}\right)$  \\ \hline
$\text{NF}$ (dB) & 48.2 & 23.3 & 9.3 & Noise figure $\left( \text{SNR}_\text{in}/\text{SNR}_\text{out} \right)$ \\ \hline
$\text{SNR}$ (dB Hz) & 135.8 & 160.7 & 174.6 & Signal to noise ratio $\left( P_\text{out}/P_\text{N}\right)$ \\ \hline
\end{tabular}
\caption{Calculated RF link properties of a two-pole PPER filter, assuming demodulation using a MZI and using the parameters from Table \ref{tbl:params}. 
Case \MakeUppercase{\romannumeral 1}: $V_{\pi} = 3.5$ V, $P^\text{(A)} = 100$ mW, 
Case \MakeUppercase{\romannumeral 2}: $V_{\pi} = 0.2$ V, $P^\text{(A)} = 100$ mW, 
Case \MakeUppercase{\romannumeral 3}: $V_{\pi} = 0.2$ V, $P^\text{(A)} = 500$ mW.}
\label{tbl:RF_fig_merit}
\end{table*}

\begin{table}[!htbp]
\centering
\renewcommand{\arraystretch}{1.3} % space between rows
\begin{tabular}{| c | c | l |} \hline
Parameter & Value & Description  \\ \hline\hline
{$P^\text{(A)}$} & 100 mW & Optical power in waveguide A  \\ \hline
$P^\text{(B)}$ & 100 mW & Optical power in waveguide B  \\ \hline
$\lambda^\text{(B)}$ & 1550 nm & Optical wavelength  \\ \hline
$\Omega_0$ & $2\pi \cdot 5$ GHz & Acoustic resonant frequency  \\ \hline
$Q$ & 1000 & Acoustic Q factor  \\ \hline
$\mu$ & $2\pi \cdot 2$ MHz & Acoustic coupling between phonons \footnote{Applies only to multi-pole filters.}  \\ \hline
$L$ & 30 mm & Length  \\ \hline
$G_\text{B}$ & 1000 $(\text{Wm})^{-1}$ & Brillouin gain  \\ \hline
$P_\text{in}^{\text{RF}}$ & 10 dBm & Input RF power  \\ \hline
{$V_{\pi}$ }& 3.5 V & Intensity modulator half-wave voltage \\ \hline
$\eta$ & 0.75 A/W & Photodiode responsivity \\ \hline
$R_\text{in}$ & 50 $\Omega$ & Input impedance  \\ \hline
$R_\text{out}$ & 50 $\Omega$ & Output impedance  \\ \hline
$H_\text{pd}$ & 0.5 & Photodiode response  \\ \hline
$\tau$ & 100 psec & MZI time delay  \\ \hline
$B_\text{RF}$ & 1 Hz & RF bandwidth  \\ \hline
$T$ & 290 K & Temperature  \\ \hline
$\gamma$ & $2\pi \cdot 5$ kHz & Laser linewidth  \\ \hline
\end{tabular}
\caption{Parameters used in the RF link analysis.}
\label{tbl:params}
\end{table}

%%%
\section{\label{sec:discussion}Discussion and conclusion}
In this work, we analyzed an RF-photonic link which utilizes a Brillouin-active PPER device, and show how this system can be used to implement an RF bandpass filter. In our analysis, we have described the output RF signal in terms of the optical, RF and acoustic parameters of the system. The fact that the RF link can be tailored through optical power, material properties, device geometry, as well the intensity modulator and photo-diode parameters, enables design flexibility and lends itself to multiple practical system schemes. 
The PPER device in the heart of this RF-photonic link provides the acousto-optic interaction that enables the RF-link to achieve narrow-band filtering, and directly affects the link performance. For example, designing PPER devices with a lower acoustic dissipation rate can yield sub-MHz filter line-shapes. Further improvements to the link performance can be achieved through mode-engineering of the optical waveguides, resulting in stronger Brillouin-gain, as well as designing devices which can support higher optical power.

The ability to design a multi-pole frequency response using a PPER device, yielding out-of-band rejection of 70 dB \cite{shin2015control}, is unique compared to other Brillouin-based RF-filtering schemes.
Additionally, an all-optical filter with a similar frequency response is challenging to realize.
An equivalent filter using two coupled ring-resonators would require each resonator to have a \textit{Q}-factor on the order of $10^8$, with precise control over the coupling rates between the rings and to the bus waveguides \cite{little1997microring}. 
Moreover, the lasers used in such a filtering scheme would need to be frequency-stabilized relative to the resonances used for the filtering operation.
Since the PPER scheme does not rely on optical resonances, it does not have these limitations, with the benefit of being optically transparent over large bandwidths \cite{shin2015control}.

Analyzing the noise sources of a PPER-based RF link, we have seen that the dominant noise is the result of thermally excited phonons, when operating at room temperature. Hence, limiting the thermal occupation of the acoustic modes by working in cryogenic temperatures or using higher frequency acoustic modes, will improve system performance. The separation of the emit and receive waveguides gives the flexibility of enhancing the link gain without adding noise to the link, resulting in a lower noise figure and larger dynamic range. This can be achieved by tailoring the properties of the emit waveguide, increasing the Brillouin coupling rate, launching higher optical power, or increasing the interaction length by designing longer devices or utilizing resonant structures \cite{Otterstrom:19}. The Brillouin noise can be further reduced by decreasing the operating temperature of the device, or by laser cooling of the mechanical modes taking part in the Brillouin process \cite{otterstrom2018optomechanical}, which may enable operation in a regime where the spontaneous Brillouin scattering is no longer the dominant noise source. Furthermore, the separation of optical tones to spatially separated waveguides reduces the effects of optical non-linearity such as four-wave mixing, which can be detrimental to filter performance. This also reduces the effect of spurious optical tones such as from unwanted reflections, degrading filter response \cite{byrnes2012photonic}, and avoids the use of circulators which may be challenging to integrate on chip.

\begin{figure}
    \centering
    \includegraphics[scale=0.7]{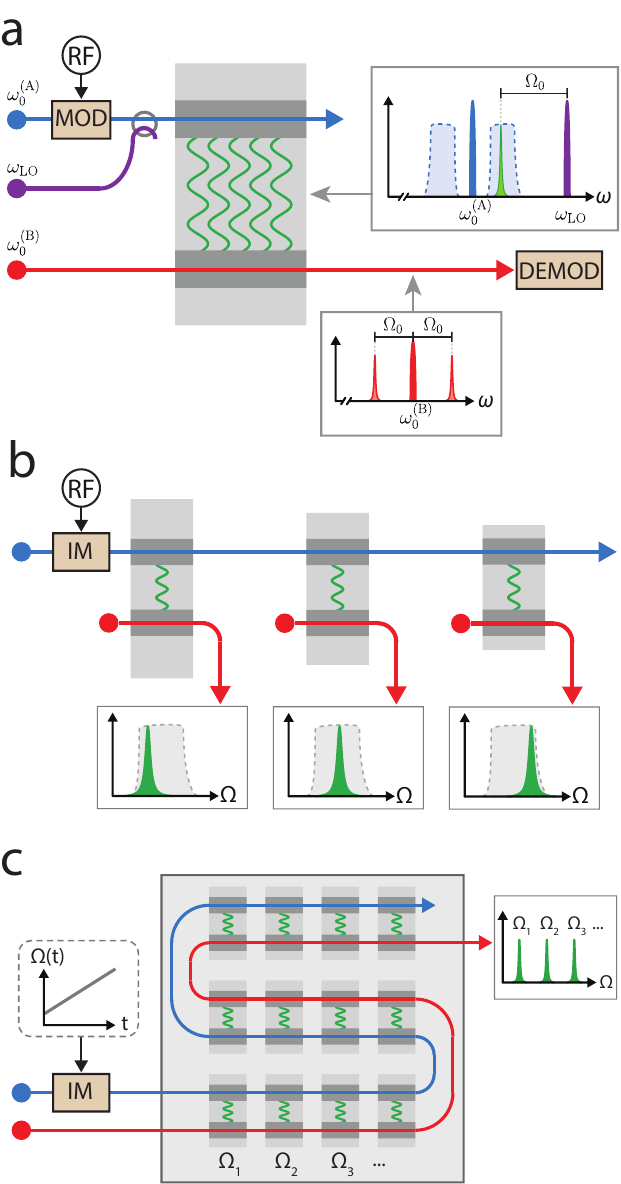}
    \caption{
    a. An RF signal is modulated on an optical carrier (blue) and combined with a second optical tone (purple). By tuning the optical wavelength, the filter passband is shifted, as the acoustic resonance (green) overlaps with different spectral bands of the modulated RF information.
    b. Cascading multiple filters in series allows to filter different frequency bands simultaneously, without the need to split the signal and with no signal degradation. 
    c. Cascading multiple PPER devices with different resonant frequencies can be used as an on-chip sensor. A swept source allows spatial resolution of the different segments, enabling distributed sensing on chip.
    }
    \label{fig:RF_applications}
\end{figure}

Since the dynamic range of this system is limited by distortion originating from the electro-optic modulator response, significant enhancement of the spur-free dynamic range can be possible with the availability of linearized intensity modulators \cite{bridges1995distortion,khilo2011broadband}, possibly yielding $\text{SFDR}_3 > 135$ dB. An important parameter for low noise figure is a low half-wave voltage of the intensity modulator at the RF link input. Using newly developed low $V_\pi$ modulators \cite{van2009low,wang2018integrated} in conjunction with higher optical powers, a noise figure of $<10$ dB becomes feasible. Alternatively, through use of a RF low-noise amplifier (LNA) to amplify the microwave signals entering the intensity modulator, this PPER filter system can yield an overall lower noise figure. For example, cascading analysis using an LNA with 30 dB of RF gain and a noise figure of 1.5 dB yields a total RF gain of 46 dB and a noise figure of 9 dB, assuming 400 mW of optical power in the emit waveguide.
The LNA in this scheme adds gain to the link without drastically changing the noise properties, resulting in a lower noise figure \cite{mahendra2018high}, and is common in the front-end design of RF receivers. However, the high gain may lead to a reduction in the dynamic range \cite{bucholtz2008graphical}, which should be considered when designing the RF link.
Additionally, biasing the intensity modulator shifted from the quadrature point can also reduce the noise figure, assuming the optical power does not change \cite{karim2007noise}.

This new type of RF-photonic link offers a variety of strategies for trimming and tuning of the bandpass frequency, and can be used in multiple frequency multiplexing schemes. The resonant frequency of the PPER can be tuned through the geometry of the device, as it sets the boundary conditions for the acoustic modes taking part in the signal transduction \cite{shin2013tailorable}. Further, by implementing different modulation schemes at the filter input, the filter bandpass frequency can be tuned optically, as illustrated in Fig. \ref{fig:RF_applications}(a). 
In this scenario, the RF information is modulated on an optical carrier (optical frequency $\omega_{0}^\text{(A)}$) using a phase modulator, and a separate optical tone at optical frequency $\omega_\text{LO}$ is used as an optical local oscillator (LO) to drive the acoustic field and transduce the information onto the light in the receive waveguide. 
As shown in the illustration in Fig. \ref{fig:RF_applications}(a), the acoustic field will be effectively driven at a frequency spaced by the Brillouin frequency ($\Omega_0$) from the LO, which in the RF domain corresponds to a pass-band around $\Omega_\text{filt} = \omega_\text{LO} - \omega_{0}^\text{(A)} - \Omega_0$.
By tuning the wavelength of the optical tones in the emit waveguide, the acoustic resonance will overlap with different spectral components of the optical sideband and filter a different spectral band. As there are no optical resonances in the device, the optical tuning of the RF filter passband can be varied over a large spectral range \cite{casas2015tunable}, while maintaining a few MHz pass-band.
The phase modulation induced by the acoustic field on the receive optical tone in waveguide B will remain around the Brillouin frequency regardless of the selected pass-band frequency, such that the demodulation and detection scheme at the link output remains unchanged from that of the static filter case. However, by utilizing different demodulation schemes at the device output, such as using the optical local oscillator to perform heterodyne detection, the RF output signal from the filter can be at same frequency as the input signal.
Furthermore, the emit and receive waveguides can operate at different wavelengths, effectively frequency shifting the optical carrier when information is transduced in the device.

Another intriguing property of the PPER RF-filtering scheme is the ability to cascade multiple filters in series without degrading the RF signal modulated on the optical field. The phonon field generated in the device phase-modulates the emit optical field as it propagates in waveguide A, but does not affect the intensity modulation that encodes the input RF signal, as discussed in Appendix \ref{subsec:app_IM_input}. Hence, the field coming out of waveguide A has an identical intensity modulation profile as the input field and will drive another PPER filter cascaded in the system exactly in the same manner \cite{OTHER_PAPER}. This is in contrast with a conventional RF filter array in which out-of-band spectral components are attenuated, leading to the need of splitting and amplifying the RF signal, resulting in lower signal power, degrading the SNR of the link, and adding to the complexity of the system. The ability to cascade multiple PPER filters in series can be of great practical use in applications such as spectral awareness and channelization \cite{umar2013comparative,anderson1991advanced}, where multiple bands of the RF spectrum are monitored simultaneously, as well as filter banks used frequently in communication systems. 
The scheme illustrated in Fig. \ref{fig:RF_applications}(b) allows us to filter out multiple spectral bands simultaneously using a single intensity modulator at the link input, and a separate receive signal-path for each spectral channel.
This can be achieved by using a separate laser source for each filtered channel, as illustrated in Fig. \ref{fig:RF_applications}(b). Alternatively, a single laser can be used, propagating through the receive waveguide of all filters and split at the output of each segment for detection.
For example, by using millimeter-scale PPER segments, each filtering a 5 MHz spectral region, 1000 filters can be cascaded, with an overall length of a few centimeters. The low optical loss that has been demonstrated in PPER devices \cite{kittlaus2018rf}, as well as the geometrically-tunable Brillouin frequency \cite{shin2013tailorable} enables this scheme to span a 5 GHz spectral range on a single chip.
Additionally, since the dominant noise source is from spontaneous Brillouin scattering, it is centered around the center frequency of each filter, and does not add out-of-band noise, as is shown in Fig. \ref{fig:RF_properties}(c), important for a cascaded filter-bank scheme.

An alternative cascading scheme is illustrated in Fig. \ref{fig:RF_applications}(c). In this scenario, a single optical receive tone propagates through multiple cascaded PPER segments, while an RF signal generator performs a frequency sweep through the range of acoustic frequencies. By designing the geometry of each PPER segment to have a different resonant frequency, the swept RF source at the input will result in multiple peaks measured on an RF spectrum analyzer, corresponding to the different segments. The high-\textit{Q} acoustic resonances, combined with the high SNR of the PPER scheme, make this an ideal candidate for on-chip sensing applications. Brillouin scattering is widely used in fiber-optic sensors, as external perturbations such as strain and temperature result in a measurable change of the acoustic resonance \cite{culverhouse1989potential,horiguchi1989tensile}. More recently, chip-scale devices implementing Brillouin scattering have been demonstrated for sensing \cite{zarifi2019chip}. Furthermore, utilizing forward-Brillouin scattering as the opto-mechanical coupling process enables interrogation of the device surface, as the transverse acoustic waves are set by the boundary conditions of the acoustic waveguide \cite{antman2016optomechanical}. This approach can be expanded to chip-scale devices, in which the device surface can be activated to bond to different chemical compounds \cite{pan2006glass,tennico2010surface}, and sensitive to nanometer-scale geometrical perturbations \cite{kittlaus2016_FSBS,kittlaus2017_Intermodal}. This enables the design of distributed sensors with micro-meter spatial resolution, and a sensitivity to perturbations, proportional to the inverse of the \textit{Q}-factor \cite{vollmer2008whispering}, of atomic-scale. Since the thermally-generated noise from each segment is centered around its resonant frequency, there is no added noise between the different cascaded sensing segments. 

While these recent demonstrations of single- and multi-pole devices were implemented in suspended silicon waveguides, the PPER concept can be realized in other systems where distinct optical fields are coupled to mutual acoustic resonances. Potential systems include multi-core fiber \cite{diamandi2017opto}, as well as  systems where optical fields are guided in different spatial modes \cite{kittlaus2018non}, or are separated in wavelength \cite{shin2013tailorable}.
The modularity of the PPER system and the large design space enable numerous applications, ranging from high performance RF photonic filters to high resolution channelizers and distributed sensors. With the advances of chip-integrated light sources, modulators and detectors, complex integrated RF-photonic systems can be feasibly engineered using the PPER scheme.
%%%

\begin{acknowledgments}
We would like to thank Yishu Zhou and Yizhi Luo for technical discussions involving Brillouin dynamics and RF-link analysis.
This research was supported primarily by the Packard Fellowship for Science and Engineering.
We also acknowledge funding support from ONR YIP (N00014-17-1-2514).
N.T.O acknowledges support from the National Science Foundation Graduate Research Fellowship under Grant DGE1122492.
Part of the research was carried out at the Jet Propulsion Laboratory, California Institute of Technology, under a contract with the National Aeronautics and Space Administration.
\end{acknowledgments}

%%%
\appendix

\section{\label{subsec:app_IM_input}Output field derivation}
We analyze the case where the input into  waveguide A is an  optical carrier at frequency \(\omega_0^\text{(A)}\) intensity modulated at frequency \(\Omega\) by an RF signal with amplitude \(V_\text{in}\). The field from a Mach-Zehnder intensity modulator (MZM) with a half-wave voltage \(V_\pi\) biased at quadrature can be described by \cite{urick2015fundamentals}
\begin{eqnarray}
    E^\text{(A)}_\text{in}(t) &&= \sqrt{P^\text{(A)}} e^{-i \omega_0^\text{(A)} t} \nonumber\\
    &&\times \left(\frac{i}{\sqrt{2}} + \frac{1}{\sqrt{2}} \exp{\left[-i \left(\pi\frac{V_\text{in}}{V_\pi}\right)  \sin{\left(\Omega t\right)}\right]}
    \right),
    \label{eq:IM_input_app}
\end{eqnarray}
or alternatively, using the Jacobi-Anger expansion can be expressed as
\begin{equation}
    E^\text{(A)}_\text{in}(t) = \sqrt{P^\text{(A)}} e^{-i \omega_0^\text{(A)} t}  \left(\frac{i}{\sqrt{2}}  + \frac{1}{\sqrt{2}} \sum_n {J_n\left(\pi\frac{V_\text{in}}{V_\pi}\right) e^{-i n\Omega t}} \right).
    \label{eq:IM_input2}
\end{equation}

Since the field is comprised of an array of optical tones spaced apart by frequency $\Omega$, we write the amplitude of each tone as $a^\text{(A)}_n(z)$ and $a^\text{(B)}_n(z)$, oscillating with frequency $\omega_0^\text{(A)} + n\Omega$ and $\omega_0^\text{(B)} + n\Omega$ for waveguides A and B respectively, and normalized using \(E^\text{}_\text{out}(t) = \sqrt{\hbar \omega_0^\text{}v} \sum_n {a}^\text{}_n e^{-i \left(\omega^\text{}_0 + n\Omega \right)t}\), such that the optical power is given by \(P^\text{(A)} = \langle |E^\text{}_\text{out} |^2 \rangle\) \cite{kharel2016_Hamiltonian}. The intensity-modulated field at the input of waveguide A, (Eq. (\ref{eq:IM_input2})) can be expressed in terms of the field amplitudes
\begin{equation}
    a_n^\text{(A)}(0) = 
    \begin{cases}
    a_0^\text{(A)} \left[\frac{i}{\sqrt{2}} + \frac{1}{\sqrt{2}}J_0\left(\pi\frac{V_\text{in}}{V_\pi}\right) \right] & n=0 \\
    a_0^\text{(A)}\frac{1}{\sqrt{2}}J_n\left(\pi\frac{V_\text{in}}{V_\pi}\right) & n\neq 0,
    \end{cases}
    \label{eq:IM_input3}
\end{equation}
and assuming a CW input into waveguide B, we have \begin{equation}
    a_n^\text{(B)}(0) = 0 \ \forall  \ n \neq 0
    \label{eq:CW_input}. 
\end{equation}

The equations of motion of the optical and acoustic field amplitudes propagating through a PPER device can by described by \cite{kang2009tightly,OTHER_PAPER}
\begin{equation}
\frac{\partial {a}^\text{(A)}_n}{\partial z} = -\frac{i}{v}\left( g b {a}^\text{(A)}_{n-1} + g^* b^{\dagger}{a}^\text{(A)}_{n+1}\right) ,
\label{eq:aA_EOM}
\end{equation}
\begin{equation}
\frac{\partial {a}^\text{(B)}_n}{\partial z} = -\frac{i}{v}\left( g b {a}^\text{(B)}_{n-1} + g^* b^{\dagger}{a}^\text{(B)}_{n+1}\right) ,
\label{eq:aB_EOM}
\end{equation}
\begin{eqnarray}
b = -i&& \left(\frac{1}{i(\Omega_0-\Omega) + \Gamma/2}\right) \nonumber\\
&& \times \sum_{n} g^* \left( {a^{\dagger}}^\text{(A)}_n {a}^\text{(A)}_{n+1} + {a^{\dagger}}^\text{(B)}_n {a}^\text{(B)}_{n+1} \right) \biggr\rvert_{z=0},
\label{eq:b_EOM}
\end{eqnarray}
where $b$ describes the acoustic field with a resonant frequency $\Omega_0$ and dissipation rate $\Gamma$, and \(v\) is the optical group velocity assumed equal in both waveguides A and B, and constant over the bandwidth in which optical scattering occurs. We  neglect optical loss in the waveguides, and the fluctuations associated with the loss of the acoustic mode will be treated in Section \ref{subsec:th_ph_noise}.

Plugging in the initial conditions from Eqs. (\ref{eq:IM_input3}) and (\ref{eq:CW_input}) into Eq. (\ref{eq:b_EOM}) gives us the phonon field
\begin{equation}
    b = - \left(\frac{1}{i(\Omega_0-\Omega)+\Gamma/2}\right) g^* \left| a_0^\text{(A)} \right|^2 J_1\left(\pi\frac{V_\text{in}}{V_\pi}\right)
    \label{eq:phonon_field_IM},
\end{equation}
and plugging this expression back into Eqs. (\ref{eq:aA_EOM}) and  (\ref{eq:aB_EOM}) yields
\begin{equation}
    \frac{\partial {a}_{n}}{\partial z}=\frac{i}{v} \left|\chi\right| \left|g\right|^2 \left| a_0^\text{(A)} \right|^2  J_1\left(\pi\frac{V_\text{in}}{V_\pi}\right) \left(a_{n-1} e^{i\phi}+a_{n+1} e^{-i\phi}\right),
    \label{eq:a_IM}
\end{equation}
for each of the waveguides A and B, where \( \chi = \left[i(\Omega_0-\Omega)+\Gamma/2\right]^{-1}\), and \(\phi = \arg{\left( \chi \right)}\). In terms of Brillouin gain and optical power we can express this as
\begin{equation}
    \frac{\partial {a}_{n}}{\partial z}= i \frac{\Gamma}{4} \left|\chi\right| G_\text{B} P^\text{(A)} J_1\left(\pi\frac{V_\text{in}}{V_\pi}\right) \left(a_{n-1} e^{i\phi} + a_{n+1} e^{-i\phi}\right),
    \label{eq:a_IM2}
\end{equation}
where the Brillouin gain corresponds to small-signal amplification such that $\partial_z P_\text{sig}(z) =  G_\text{B} P_\text{p} P_\text{sig}(z)$. The gain has units of (Power$\times$Length)$^{-1}$, and can be expressed in terms of the coupling rate as \(G_\text{B} = 4 \left|g\right|^2 / \left(\hbar \omega v^2 \Gamma \right)\) \cite{kharel2016_Hamiltonian}.

\begin{figure}
    \centering
    \includegraphics[scale=0.7]{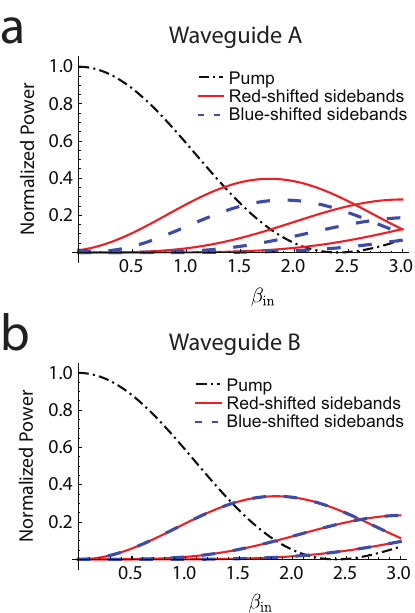}
    \caption{
    Normalized power of the optical tones, given an intensity modulated input to waveguide A, and a single tone input into waveguide B. In each waveguide, the pump tone is denoted in black, Stokes (red-shifted) tones in red, and anti-Stokes (blue-shifted) tones are shown in blue.
    }
    \label{fig:Bessel}
\end{figure}
The recurrence relation obtained in Eq. (\ref{eq:a_IM2}) is consistent with a modified Bessel function following \( I_n' = \frac{1}{2} \left( I_{n-1} + I_{n+1} \right)\), such that the optical fields can be written as a linear combination \(a_n(z) = e^{i\phi n} \sum_m c_{n,m} I_m \left(i \left(\Gamma/2\right) \left|\chi\right| G_\text{B} P^\text{(A)}  J_1\left(\pi V_\text{in}/V_\pi\right) z\right)\). We can find the coefficients \(c_{n,m}\) by using the identity \(I_m(0) = \delta_{m,0} \), such that \(c_{n,m} = a_{n-m}(0) e^{-i\phi (n-m)} \), and by using the relation \( I_m(x) = i^{-m}J_m(ix)\) we have
\begin{equation}
    a_n(z) = \sum_m a_{n+m}(0) i^{m} J_{m} \left( \beta_\text{in} \right)e^{-i\phi m},
    \label{eq:a5}
\end{equation}
for both waveguides A and B, where \(\beta_\text{in} = (\Gamma/2) \left|\chi\right| G_\text{B} P^\text{(A)} J_1\left(\pi V_\text{in}/V_\pi\right) z \). We can now plug in the initial conditions from Eq. (\ref{eq:CW_input}) to calculate the field in waveguide B
\begin{equation}
    a_n^\text{(B)}(z) = a_{0}^\text{(B)}(0) i^{-n} J_{-n} \left( \beta_\text{in} \right) e^{i \phi n}.
    \label{eq:a6}
\end{equation}
The power of the different optical tones in the two waveguides as a function of $\beta_\text{in}$, following Eq. (\ref{eq:a5}) and the initial conditions for each waveguide is shown in Fig. \ref{fig:Bessel}.

Summing all the mode amplitudes in waveguide B yields \(E^\text{(B)}_\text{out}(t) = \sqrt{\hbar \omega_0^\text{(B)}v} \sum_n {a}^\text{(B)}_n e^{-i \left(\omega^\text{(B)}_0 + n\Omega \right)t}\), and using the Jacobi-Anger expansion, \(\sum_n i^n J_n(z) e^{ixn} = e^{iz\cos{x}}\), we can rewrite the field in waveguide B as
\begin{equation}
    E^\text{(B)}_\text{out}(t) = \sqrt{P^\text{(B)}} \ e^{-i \omega^\text{(B)}_0t} \ \exp \left[i\ \beta_\text{in}  \cos{\left(\Omega t - \phi\right)} \right],
    \label{eq:waveguide_B_output_IM_app}
\end{equation} 
revealing the phase modulation that the tone in waveguide B experiences as it propagates through the device, discussed further in Ref. \cite{OTHER_PAPER}.

We can also analyze the field coming out of waveguide A by plugging in the input mode amplitudes from Eq. (\ref{eq:IM_input3}) into Eq. (\ref{eq:a5}), giving us
\begin{eqnarray}
    a_n^\text{(A)}(z) =&& a_0^\text{(A)} \sum_m i^{m} J_{m} \left( \beta_\text{in} \right)e^{-i\phi m} \nonumber \\
    && \times \left[\frac{1}{\sqrt{2}} J_{n+m} \left(\pi\frac{V_\text{in}}{V_\pi}\right) + \frac{i}{\sqrt{2}} \delta_{n+m,0} \right].
\end{eqnarray}

Summing all the mode amplitudes, $E^\text{(A)}_\text{out}(t) = \sqrt{\hbar \omega_0^\text{(A)}v} \sum_n {a}^\text{(A)}_n e^{-i \left(\omega^\text{(A)}_0 + n\Omega \right)t}$, and using a similar derivation to the one we had for waveguide B, we arrive at
\begin{eqnarray}
    E^\text{(A)}_\text{out}(t) &&= \sqrt{P^\text{(A)}} \ e^{-i \omega^\text{(A)}_0t} \ \exp \left[i\ \beta_\text{in}  \cos{\left(\Omega t - \phi\right)} \right] \nonumber \\ 
    &&\times \sum_n e^{-i n\Omega t} \left[\frac{1}{\sqrt{2}} J_{n} \left(\pi\frac{V_\text{in}}{V_\pi}\right) + \frac{i}{\sqrt{2}} \delta_{n,0} \right],
    \label{eq:waveguide_A_output_IM_app1}
\end{eqnarray} 
and by using Eq. ({\ref{eq:IM_input2}}), this can also be written as
\begin{eqnarray}
    E^\text{(A)}_\text{out}(t) = E^\text{(A)}_\text{in}(t) \ \exp \left[i\ \beta_\text{in} \cos{\left(\Omega t - \phi\right)} \right],
    \label{eq:waveguide_A_output_IM_app}
\end{eqnarray} 
showing that the input undergoes pure phase-modulation. Calculating the beat-note at frequency $\Omega$ reveals that it is unchanged as the field propagates in waveguide A, i.e. $\sum_n {a_n^\text{(A)}}^\dagger(z) \ a_{n+1}^\text{(A)} (z) = |a_0^\text{(A)}|^2 J_1(\pi V_\text{in} / V_\pi)$.

\section{\label{subsec:app_bal_demod}Balanced-detection demodulation}
\begin{figure}[bt!]
\centering
    \includegraphics[scale=0.7]{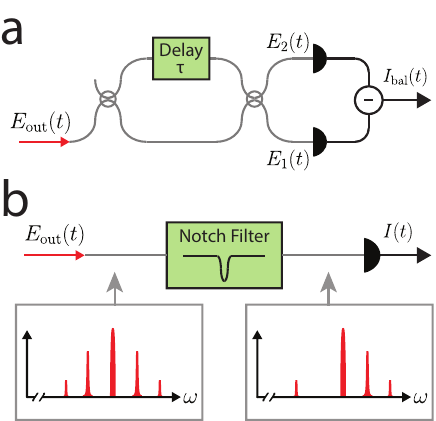}
    \caption{a. An unbalanced MZI splits the input field, and adds a phase shift corresponding to a time delay \(\tau\) to one of the two arms. The two interferometer arms are then combined using another coupler and detected using two photodiodes. The current of the two diodes is subtracted from each other for balanced detection. b. A different phase  demodulation scheme, using a narrow-band notch filter to eliminate one of the optical sidebands.}
    \label{fig:demod}
\end{figure}

The information in the form of phase modulation of the light at the output of waveguide B needs to be demodulated back into the RF domain. We assume a phase demodulator consisting of a Mach-Zehnder interferometer (MZI), illustrated schematically in Fig. \ref{fig:demod}(a), composed of a $50:50$ directional coupler, a time delay on one of the two interferometer arms, and another $50:50$ coupler. This can be described in terms of matrix operations on the input field amplitudes \cite{urick2015fundamentals}
\begin{equation}
    \begin{pmatrix} E_1(t) \\ E_2(t) \end{pmatrix} = \frac{1}{2}
    \begin{pmatrix} 1 & i \\ i & 1 \end{pmatrix}
    \begin{pmatrix} \hat{\Gamma}\left(\tau\right) & 0 \\ 0 & 1 \end{pmatrix}
    \begin{pmatrix} 1 & i \\ i & 1 \end{pmatrix}
    \begin{pmatrix} E_\text{out}(t) \\ 0 \end{pmatrix},
\end{equation}
where we assume ideal lossless couplers, and \(\hat{\Gamma}\left(\tau\right)\) is a time delay operator such that \( \hat{\Gamma}\left(\tau\right) E(t) = E(t-\tau)\). The input to the MZI is the field from the output of waveguide B, denoted $E_\text{out}(t)$. This yields the fields at the output of the MZI 
\begin{equation}
    \begin{split}
    E_1 = \frac{1}{2}\Big(E_\text{out}(t-\tau) - E_\text{out}(t)\Big),\\
    E_2 = \frac{i}{2}\Big(E_\text{out}(t-\tau) + E_\text{out}(t)\Big),
    \end{split}
    \label{eq:bal_fields}
\end{equation}
which are each directed at a detector, as illustrated in Fig. \ref{fig:demod}(a). The photo-current generated in each of the photo-diodes \(I(t) = \eta |E(t)|^2\) is subtracted, such that \(I_\text{bal}(t) = \eta \left( |E_2(t)|^2 - |E_1(t)|^2\right)\), assuming the two detectors have equal responsivities \(\eta\). Plugging in Eq. (\ref{eq:bal_fields}) gives us
\begin{eqnarray}
    I_\text{bal}(t) &&= \frac{1}{2} \eta \Big( E_\text{out}^*(t-\tau)E_\text{out}(t) + E_\text{out}(t-\tau)E_\text{out}^*(t) \Big) \nonumber\\
    &&= \eta \ \Re \Big(E_\text{out}^*(t-\tau)E_\text{out}(t) \Big),
    \label{eq:I_bal}
\end{eqnarray}
and plugging in the output field from the device, obtained in Eq. (\ref{eq:waveguide_B_output_IM_app}), yields
\begin{eqnarray}
    I_\text{bal}(t) = && \eta P^\text{(B)}\Re \Big( e^{-i\omega_0\tau}  \Big. \nonumber \\
    && \times \Big. \exp \Big[i \ \beta_\text{in} \Big(\cos{(\Omega t-\phi)} - \cos{\left(\Omega(t-\tau)-\phi\right)}\Big)\Big]\Big), \nonumber\\
\end{eqnarray}
where we are using the notation \(\beta_\text{in} = (\Gamma/2) |\chi| J_1(\pi V_\text{in}/V_\pi) G_\text{B} P^\text{(A)} L \). Using trigonometric and Bessel function properties, this can be rewritten as
\begin{eqnarray}
I_\text{bal}(t) = && \eta P^\text{(B)} \Re \Big( e^{-i\omega_0\tau} \Big. \nonumber \\
&&\times \Big.\sum_n J_n \Big(2\beta_\text{in}\sin{(\Omega \tau/2)}\Big) e^{-in(\Omega t-\Omega \tau/2-\phi)} \Big).
\end{eqnarray}

We can isolate the term oscillating at frequency \(\Omega\) by keeping only the \(n= \pm 1\) terms in the sum and using the Bessel function property \(J_1(x) = -J_{-1}(x)\), giving us
\begin{eqnarray}
I_\text{bal}^\Omega(t) = \eta P^\text{(B)} J_1 && \Big(2\beta_\text{in}\sin{(\Omega \tau/2)}\Big) \nonumber \\
\times && \Big ( \cos{(\Omega t - \Omega \tau/2 - \phi+\omega_0\tau)} \Big. \nonumber\\
&& - \Big. \cos{(\Omega t-\Omega \tau/2-\phi -\omega_0\tau)}\Big),
\end{eqnarray}
which can also be expressed as
\begin{eqnarray}
    I_\text{bal}^\Omega(t) = -2 \eta && P^\text{(B)} J_1 \Big( 2\beta_\text{in}\sin{(\Omega \tau/2)} \Big) \nonumber \\ &&\times \sin{(\Omega t-\Omega \tau/2 - \phi)} \sin{(\omega_0\tau)}.
\end{eqnarray}
We will assume the MZI is operating at quadrature, such that \(\omega_0\tau = \pi (m+1/2) \) for an integer \(m\). The RF power at frequency \(\Omega\) is now given by
\begin{eqnarray}
    P_\text{out}^\text{RF}(\Omega) =&& 2 R_\text{out} |H_\text{pd}|^2 \nonumber \\
    && \times \Bigg(\eta P^\text{(B)} J_1 \Big(2\beta_\text{in}\sin{(\Omega \tau/2)}\Big) \Bigg)^2,
    \label{eq:P_out_RF2}
\end{eqnarray}
where we have used \(P_\text{out}^\text{RF} =\langle I_\Omega^2\rangle R_\text{out} |H_\text{pd}|^2\), and $R_\text{out}$, $H_\text{pd}$  denote the output impedance and the photodiode circuit efficiency, respectively \cite{urick2015fundamentals}. Assuming the input intensity modulation is in the small-signal regime, we can linearize the Bessel function \(J_1\left(\pi V_\text{in}/V_\pi\right) \approx \pi V_\text{in}/\left(2 V_\pi\right) \), such that \(\beta_\text{in} \approx \left(\Gamma/4\right) \left|\chi\right| \left(\pi V_\text{in}/V_\pi\right) G_\text{B} P^\text{(A)} L\), yielding
\begin{eqnarray}
   &&P_\text{out}^\text{RF}(\Omega) = 2 R_\text{out} |H_\text{pd}|^2 \nonumber \\ 
   && \times \Bigg(\eta P^\text{(B)} J_1 \left( \frac{\Gamma}{2} \left|\chi\right| \frac{\pi V_\text{in}}{V_{\pi}} G_\text{B} P^\text{(A)} L \sin{(\Omega \tau/2)}\right) \Bigg)^2 .
    \label{eq:P_out_RF_app}
\end{eqnarray}

\section{\label{subsec:th_ph_noise}Balanced-detection thermal-phonon noise}
When an optical tone propagates through a Brillouin-active region, the thermally occupied phonon population results in light scattering to sidebands spaced $\Omega_0$ around the optical frequency $\omega_0$ \cite{kharel2016_Hamiltonian}. Assuming a single tone input into the waveguide, the optical field at the output can be described by
\begin{eqnarray}
    E_\text{out}&&(t) = \sqrt{\hbar \omega_0 v} \nonumber \\
    && \times \left( a_{-1} e^{-i (\omega_0-\Omega_0) t} + a_0 e^{-i \omega_0 t} + a_{1} e^{-i (\omega_0+\Omega_0) t} \right),
    \label{eq:noise_EOM}
\end{eqnarray}
where the amplitudes $a_{-1}$ and $a_{1}$ are the sidebands generated by the spontaneous scattering, given by \cite{raymer1981stimulated,boyd1990noise,kharel2016_Hamiltonian}
\begin{equation}
    \begin{split}
        a_{-1}(z,\tau) &= -i\frac{g^*}{v}a_{0} \int_0^\tau d\tau'\int_0^z dz'\eta^\dagger(z',\tau') e^{-\frac{\Gamma}{2}(\tau-\tau')},\\
        a_{1}(z,\tau) &= -i\frac{g}{v}a_{0} \int_0^\tau d\tau'\int_0^z dz'\eta(z',\tau') e^{-\frac{\Gamma}{2}(\tau-\tau')},
    \end{split}
    \label{eq:noise_sidebands}
\end{equation}
where \(\eta(z,t)\) is the Langevin force corresponding to the phonon dissipation, such that \(\langle \eta(z,t)\rangle = 0\) and \(\langle \eta^\dagger(z,t)\eta(z',t')\rangle = \bar{n}_\text{th}\Gamma \delta(z-z')\delta(t-t')\), where $\bar{n}_\text{th}$ is the mean thermal phonon occupation number. At room temperature, when $\bar{n}_\text{th} \gg 1 $, this thermal occupation  can be approximated using  $\bar{n}_\text{th} \approx (k_\text{B} T)/(\hbar \Omega_0)$.

For the MZI balanced-detection scheme described earlier, we can calculate the photo-current by plugging Eqs. (\ref{eq:noise_EOM}) and (\ref{eq:noise_sidebands}) into Eq. (\ref{eq:I_bal}). The noise power spectral density can then be calculated using the Wiener-Khinchin theorem,
$
    S_\text{N}^\text{RF}(\Omega) = R_\text{out} |H_\text{pd}|^2 \int dt' e^{-i \Omega t'}\langle I_\text{N}(t+t') I_\text{N}(t) \rangle,
$
and keeping terms oscillating around frequency \(\Omega_0\) yields
\begin{widetext}
\begin{equation}
    \begin{split}
    {S_\text{N}}_\text{bal}^\text{RF}(\Omega) = 4 \eta^2 \frac{\omega_0}{\Omega_0} G_\text{B} P^2 L k_\text{B} T R_\text{out} |H_\text{pd}|^2 \Bigg[ \sin^2{\left( \Omega \tau /2 \right)} \Bigg(&\frac{(\Gamma/2)^2}{(\Gamma/2)^2 + (\Omega-\Omega_0)^2} + \frac{(\Gamma/2)^2}{(\Gamma/2)^2 + (\Omega+\Omega_0)^2}\Bigg) \\
    & -\frac{\Omega_0 \Gamma  e^{-(\Gamma/2)\tau} \left( (\Gamma/2)^2 - \Omega^2 + \Omega_0^2\right) \sin{(\Omega_0\tau)} }{\left[(\Gamma/2)^2 + (\Omega-\Omega_0)^2\right]\left[(\Gamma/2)^2 + (\Omega+\Omega_0)^2\right]}
    \Bigg],
    \end{split}
    \label{eq:bal_noise_full}
\end{equation}
\end{widetext}
where we have assumed the MZI is operating at quadrature, such that $e^{-2i\omega_0 \tau} =-1$.

We now set the time delay for maximum transmission on resonance, such that \(\Omega_0\tau/2 = \pi (m+1/2) \), yielding
\begin{eqnarray}
    &&{S_\text{N}}_\text{bal}^\text{RF}(\Omega) = 4 \eta^2 \frac{\omega_0}{\Omega_0} G_\text{B} P^2 L k_\text{B} T R_\text{out} |H_\text{pd}|^2 \sin^2{\left( \Omega \tau /2 \right)} \nonumber\\
    && \times \Bigg[ \frac{(\Gamma/2)^2}{(\Gamma/2)^2 + (\Omega-\Omega_0)^2} + \frac{(\Gamma/2)^2}{(\Gamma/2)^2 + (\Omega+\Omega_0)^2} \Bigg].
    \label{eq:SN_bal_app}
\end{eqnarray}

For a narrow RF bandwidth \(B_\text{RF}\) centered around \(\Omega\), the integrated power will be
\begin{eqnarray}
    {P_\text{N}}_\text{bal} = 8 \eta^2 \frac{\omega_0}{\Omega_0} G_\text{B} P^2 L k_\text{B} T && B_\text{RF} R_\text{out} |H_\text{pd}|^2 \sin^2{\left( \Omega \tau /2 \right)} \nonumber\\
    &&\times \frac{(\Gamma/2)^2}{(\Gamma/2)^2 + (\Omega-\Omega_0)^2},
\end{eqnarray}
where \(B_\text{RF}\) is in units of Hz, and we have re-written the power as a single-sideband spectrum.

\section{\label{subsec:filt_demod}Sideband filtering demodulation}
We describe another demodulation scheme, where the first sideband of the phase modulated output at frequency \(\omega_0^\text{(B)} - \Omega\) is optically filtered to turn the phase fluctuations into an intensity signal that can be detected using a single photodetector. This scheme was implemented in Ref. \cite{kittlaus2018rf} and is illustrated in Fig. \ref{fig:demod}(b).
The output field is now given by \(E^\text{(B)}_\text{out}(t) = \sqrt{\hbar \omega_0^\text{(B)}v} \sum_{n \neq -1} {a}^\text{(B)}_n e^{-i \left(\omega^\text{(B)}_0 + n\Omega \right)t}\), and  directing the output light to a photodiode will result in a photo-current \(I = \eta |E^\text{(B)}_\text{out}(t)|^2\) where \(\eta\) is the responsivity of the photodiode. Plugging in Eq. (\ref{eq:waveguide_B_output_IM_app}) and keeping terms oscillating around frequency \(\Omega\) yields
\begin{eqnarray}
 I_\Omega = 2 \eta P^\text{(B)} && \cos{\left(\Omega t+\pi/2 - \phi\right)} \nonumber \\
 &&\times \left[ J_0\left( \beta_\text{in} \right)J_1\left( \beta_\text{in} \right) - J_1\left( \beta_\text{in} \right)J_2\left( \beta_\text{in} \right) \right].
\end{eqnarray}

The RF output average power at frequency \(\Omega\), given by \(P_\text{out}^\text{RF} =\langle I_\Omega^2\rangle R_\text{out} |H_\text{pd}|^2\) where again \(H_\text{pd}\) is the photodiode circuit efficiency and \(R_\text{out}\) the output impedance \cite{urick2015fundamentals}, is now
\begin{eqnarray}
P_\text{out}^\text{RF}&&(\Omega) = 2 R_\text{out} |H_\text{pd}|^2 \nonumber \\
&& \times \Big(\eta P^\text{(B)}\left[ J_0\left( \beta_\text{in} \right)J_1\left( \beta_\text{in} \right) - J_1\left( \beta_\text{in} \right)J_2\left( \beta_\text{in} \right) \right]\Big)^2.
\label{eq:RF_Pout}
\end{eqnarray}

Assuming the input intensity modulator is in the small-signal linear regime, we can linearize the Bessel function in \(\beta_\text{in}\) to first order \(\beta_\text{in} \approx \left(\Gamma/4\right) \left|\chi\right| \left(\pi V_\text{in}/V_{\pi}\right)G_\text{B} P^\text{(A)} L\), and drop the second-order Bessel function, yielding
\begin{widetext}
\begin{equation}
P_\text{out}^\text{RF}(\Omega) = 2\Bigg(\eta P^\text{(B)} J_0\left( \frac{\Gamma}{2} \left|\chi\right| \frac{\pi}{2}\frac{V_\text{in}}{V_{\pi}} G_\text{B} P^\text{(A)} L \right)J_1\left( \frac{\Gamma}{2} \left|\chi\right| \frac{\pi}{2}\frac{V_\text{in}}{V_{\pi}} G_\text{B} P^\text{(A)} L \right) \Bigg)^2 R_\text{out} |H_\text{pd}|^2.
\label{eq:P_RF_out_ss}
\end{equation}
\end{widetext}

\subsection*{\label{subsec:filt_demod_noise}Sideband filtering demodulation thermal-phonon noise}
After filtering out one of the sidebands in the demodulation process, the field describing the spontaneous scattered light (Eq. \ref{eq:noise_EOM}) has only one sideband
\begin{equation}
    E_\text{out}(t) = \sqrt{\hbar \omega_0 v} \left(a_0 e^{-i \omega_0 t} + a_{1} e^{-i (\omega_0+\Omega_0) t} \right).
    \label{eq:noise_EOM_filtered}
\end{equation}

Calculating the photocurrent generated by this spontaneous scattering \(I_\text{N}(t) = \eta \left|E_\text{out}(t)\right|^2\) gives us
\begin{equation}
    I_\text{N}(t) = \eta \hbar \omega_0 v \left(  |a_0|^2 + |a_{1}|^2 + 2\Re{ \left(a_0^\dagger a_1\right) \cos{\left(\Omega_0 t\right)}} \right).
\end{equation}

Keeping terms oscillating around \(\Omega_0\) and calculating the spectral density using 
the Wiener-Khinchin theorem results in
\begin{eqnarray}
    &&S_\text{N}^\text{RF}(\Omega) = \eta^2 \frac{\omega_0}{\Omega_0} G_\text{B} P^2 L k_\text{B} T R_\text{out} |H_\text{pd}|^2 \nonumber\\
    &&\times\Bigg[ \frac{(\Gamma/2)^2}{(\Gamma/2)^2 + (\Omega-\Omega_0)^2} + \frac{(\Gamma/2)^2}{(\Gamma/2)^2 + (\Omega+\Omega_0)^2} \Bigg],
    \label{eq:SN_filt}
\end{eqnarray}
and integrating a narrow bandwidth \(B_\text{RF}\) around frequency \(\Omega\), the single-sideband power spectral density is
\begin{eqnarray}
    P_\text{N}(\Omega) = 2 \eta^2 \frac{\omega_0}{\Omega_0} G_\text{B} P^2 L && k_\text{B} T B_\text{RF} R_\text{out} |H_\text{pd}|^2 \nonumber\\
    &&\times \frac{(\Gamma/2)^2}{(\Gamma/2)^2 + (\Omega-\Omega_0)^2}.
    \label{eq:PN_filt}
\end{eqnarray}

In terms of RIN \cite{urick2015fundamentals}, this can be expressed as
\begin{equation}
    \text{RIN}_\text{phonons} = 2 \frac{\omega_0}{\Omega_0} G_\text{B} L k_\text{B} T \frac{(\Gamma/2)^2}{(\Gamma/2)^2+(\Omega - \Omega_0)^2},
\end{equation}
which is four times lower than the balanced detection result shown earlier in Eq. \ref{eq:thermal_ph_bal}.
\vspace{5mm} 

The filtering of the optical sideband turns the phase fluctuations into intensity, resulting in the laser phase noise being converted into intensity noise, with a RIN given by 
\begin{eqnarray}
    &&\text{RIN}_\text{phase} = \nonumber\\ &&\frac{2}{\pi}\int_{-B/2}^{B/2}{d\Omega' \left(\frac{\gamma}{\left(\gamma/2\right)^2 + \left(\Omega'\right)^2}\right) \left(\frac{\gamma}{\left(\gamma/2\right)^2 + \left(\Omega' - \Omega\right)^2}\right) }, \nonumber\\
\end{eqnarray}
where \(B\) is the bandwidth of the optical filter used to reject one of the sidebands, and $\gamma$ is the laser linewidth resulting from its phase fluctuations. As with the MZI demodulation case, using typical values, the thermal phonon scattering contribution is still the dominant noise source \cite{kittlaus2018rf}.

\subsection*{\label{subsec:filt_link}Sideband filtering demodulation RF-link properties}
In the small-signal limit we can expand the Bessel functions to first order, \(J_0(x) \approx 1\), \(J_1(x) \approx x/2\). Writing in terms of input RF average power \cite{urick2015fundamentals}  \(P_\text{in}^{\text{RF}}=V_{\text{in}}^2/\left(2R_{\text{in}}\right)\), where \( R_{\text{in}} \) is the input impedance, yields
\begin{equation}
P_\text{out}^\text{RF}(\Omega) = \frac{1}{4} P_\text{in}^{\text{RF}}\left(\eta P^\text{(B)} \frac{\Gamma}{2} \left|\chi\right| \frac{\pi}{V_{\pi}} G_\text{B} P^\text{(A)} L\right)^2 R_\text{in} R_\text{out} |H_\text{pd}|^2.
\label{eq:P_RF_out_ss_lin}
\end{equation}

We observe that the RF link gain \(g = P_\text{out}^\text{RF}/P_\text{in}^{\text{RF}}\) is four times smaller than obtained by the interferometric demodulation approach described in the previous sections, shown in  Eq. (\ref{eq:P_out_RF_lin2}). This is consistent with the calculation in the previous section, also showing a four-fold reduction in the noise power, resulting in the same signal-to-noise ratio for both demodulation schemes.

Next, we calculate the \(1 \ \text{dB} \) compression point, i.e. the input RF power at which the output is 1 dB lower than the linearized response. Setting the ratio between Eqs. (\ref{eq:P_RF_out_ss}) and (\ref{eq:P_RF_out_ss_lin}) to 1 dB, we solve numerically to find the input power where the output is compressed by 1 dB, yielding
\begin{equation}
    P_\text{in}^{\text{1dB}} =\frac{0.245}{R_\text{in}} \left(\frac{V_{\pi}}{\Gamma \left|\chi\right| G_\text{B} P^\text{(A)} L}\right)^2.
\end{equation}

We now calculate the third-order intercept point \cite{urick2015fundamentals}. We assume the input RF signal is at frequency \(\Omega_0/3\) and see the propagation of the third-order modulation through the device. In this case we have \(\beta_\text{in}^{3\Omega/3} = \left(\Gamma/2\right) \left|\chi\right| J_3\left(\pi V_\text{in}/V_{\pi}\right)G_\text{B} P_0^{(A)} L\), and plugging into Eq. (\ref{eq:RF_Pout}), expanding the Bessel functions to first order yields
\begin{equation}
P_\text{out}^{3\Omega/3} = \frac{1}{2}\Big(\eta P^\text{(B)} \frac{\Gamma}{2} \left|\chi\right| J_3\left(\pi\frac{V_\text{in}}{V_{\pi}}\right) G_\text{B} P^\text{(A)} L \Big)^2 R_\text{out} |H_\text{pd}|^2.
\end{equation}

Expanding \(J_3\left(\pi V_\text{in}/V_{\pi}\right) \) to the first non-vanishing order, \(J_3(x) \approx x^3/48\), leaves us with
\begin{eqnarray}
P_\text{out}^{3\Omega/3} &&=\frac{1}{576} R_\text{in}^3 R_\text{out} |H_\text{pd}|^2 \nonumber\\
&&\times \left (P_\text{in}^{\text{RF}}\right)^3 \Big(\eta P^\text{(B)} \frac{\Gamma}{2} \left|\chi\right| \left(\frac{\pi}{V_{\pi}}\right)^3 G_\text{B} P^\text{(A)} L \Big)^2.
\label{eq:P_RF_out_3Omega_ss}
\end{eqnarray}

To find the third-order intercept point, we equate Eqs. (\ref{eq:P_RF_out_ss_lin}) and (\ref{eq:P_RF_out_3Omega_ss}), and solve for the input RF power, giving the input intercept point ($\text{IIP}_3$)
\begin{equation}
    \text{IIP}_3 = 12\frac{V_{\pi}^2}{\pi^2 R_\text{in}},
\end{equation}
and plugging back into Eq. (\ref{eq:P_RF_out_ss_lin}) gives us the output intercept point ($\text{OIP}_3$)
\begin{equation}
    \text{OIP}_3 = 3 \left(\eta P^\text{(B)} \frac{\Gamma}{2} \left|\chi\right| G_\text{B} P^\text{(A)} L \right)^2R_\text{out} |H_\text{pd}|^2.
\end{equation}

As we have shown, the signal and the noise both change by the same factor using the alternate demodulation scheme, and the noise figure and spur-free dynamic range will not change. However, different demodulation schemes may have different signal compression, yielding a difference in the linear dynamic range.

\bibliography{RF_PPER}

\end{document}